\documentclass[11pt,a4paper]{article}
\usepackage{amsmath,amssymb}
\usepackage[pdftex]{graphicx}
\usepackage[version=3]{mhchem}
\usepackage{authblk}
\usepackage{multicol}
\usepackage{float}
\usepackage{indentfirst}
\usepackage{fancyhdr}

  \renewcommand{\footrulewidth}{0.4pt}

\makeatletter

\def\@thesis{Talk presented at the International Workshop on Future Linear Colliders (LCWS2018), Arlington, Texas, 22-26 October 2018. C18-10-22.}
\def\author#1{\def\@author{#1}}
\def\department#1{\def\@department{#1}}

\def\@maketitle{
\begin{center}
\hrule height 0.5mm depth 0mm width 166mm
\vspace{5mm}
{\Large \@thesis \par} 
\vspace{5mm}
\hrule height 0.5mm depth 0mm width 166mm
\vspace{10mm}
{\Large\bf \@title \par}
\vspace{10mm}
{\large \@author \par}
\vspace{3mm}
{\large \@department}
\end{center}
\vspace{3mm}
}

\makeatother

\title{Performance of alternative scintillator tile geometry for AHCAL}
\author{Naoki Tsuji, on behalf of the CALICE Collaboration}
\department{The University of Tokyo, 7-3-1, Hongo, Bunkyo-ku, Tokyo 113-0033, Japan}
\date{\empty}

\begin{document}

\maketitle

\begin{abstract}
The CALICE Analogue Hadron CALorimeter (AHCAL) at the International Linear Collider (ILC) is a high-granularity hadron calorimeter based on scintillator tiles readout by MPPCs. Toward the construction of ILC, the optimization of the AHCAL granularity is
revisited, and we are studying mixed granularity with larger scintillator tile. We first measured the performance of $60\times60~\mathrm{mm^2}$ tile, which is larger than the standard $30\times30~\mathrm{mm^2}$ tile. The light yield of $60\times60~\mathrm{mm^2}$ tile was measured to be about a half of that of $30\times30~\mathrm{mm^2}$ tile, while the uniformity of the tile response turned out to be very good. Then, a detection layer composed of 144 tiles of $60\times60~\mathrm{mm^2}$ was constructed. The detection layer was added to the large technological prototype of AHCAL composed of 38 detection layers with $30\times30~\mathrm{mm^2}$ tiles. The detection layer was successfully tested in test beam experiments at CERN SPS.
\end{abstract}

\begin{multicols}{2}

\section{Introduction}
\pagestyle{fancy}
  \cfoot{LCWS2018}
  \rfoot{\thepage}
  \renewcommand{\footrulewidth}{0.4pt}

The International Linear Collider (ILC) is a future electron-positron linear collider with a center-of-mass energy 250--500~GeV, which is extendable to 1~TeV. The International Large Detector (ILD), one of the two detector concepts for ILC, consists of a vertex detector, a central tracker, an EM calorimeter, a hadron calorimeter, and a muon detector. ILD adopts the Particle Flow Algorithm (PFA), which reconstructs individual particles using the best suited detectors to significantly improve the jet energy resolution. In order to realize PFA, high-granularity calorimeter is required.

The Analogue Hadron CALorimeter (AHCAL) is developed in the framework of the CALICE collaboration\cite{Sefkow:2018rhp}. It consists of 48 layers (each of which is a combination of a scintillator active layer and a steel or tungsten absorber layer) with 8 million pieces of scintillator tiles (30$\times$30$\times$3$~\mathrm{mm^3}$ each) readout by a MPPC (1.3$\times$1.3$~\mathrm{mm^2}$). The AHCAL group constructed a large technological prototype, which consists of 38 layers. Each layer is based on four HCAL Base Units (HBUs), each of which has 144 tiles of 30$\times$30$~\mathrm{mm^2}$. It was tested in test beam experiments at CERN SPS.

Toward the construction of ILC, it is necessary to make a more realistic design of the ILD detector. The granularity of AHCAL is now revisited. We study mixed granularity configurations using different sizes of tiles at different layers. In these configurations, the number of readout channels and ASICs, thus the cost can be reduced. According to a previous study\cite{previous}, 30$\times$30$~\mathrm{mm^2}$ tiles can be replaced with 60$\times$60$~\mathrm{mm^2}$ tiles up to latter half of the layers without significant performance degradation. However, the performance of tiles larger than 30$\times$30$~\mathrm{mm^2}$ has not ever been tested.

\section{Performance of 60$\times$60$~\mathrm{mm^2}$ tile}
\begin{figure}[H]
	\centering
	\includegraphics[width=80mm,pagebox=cropbox,clip]{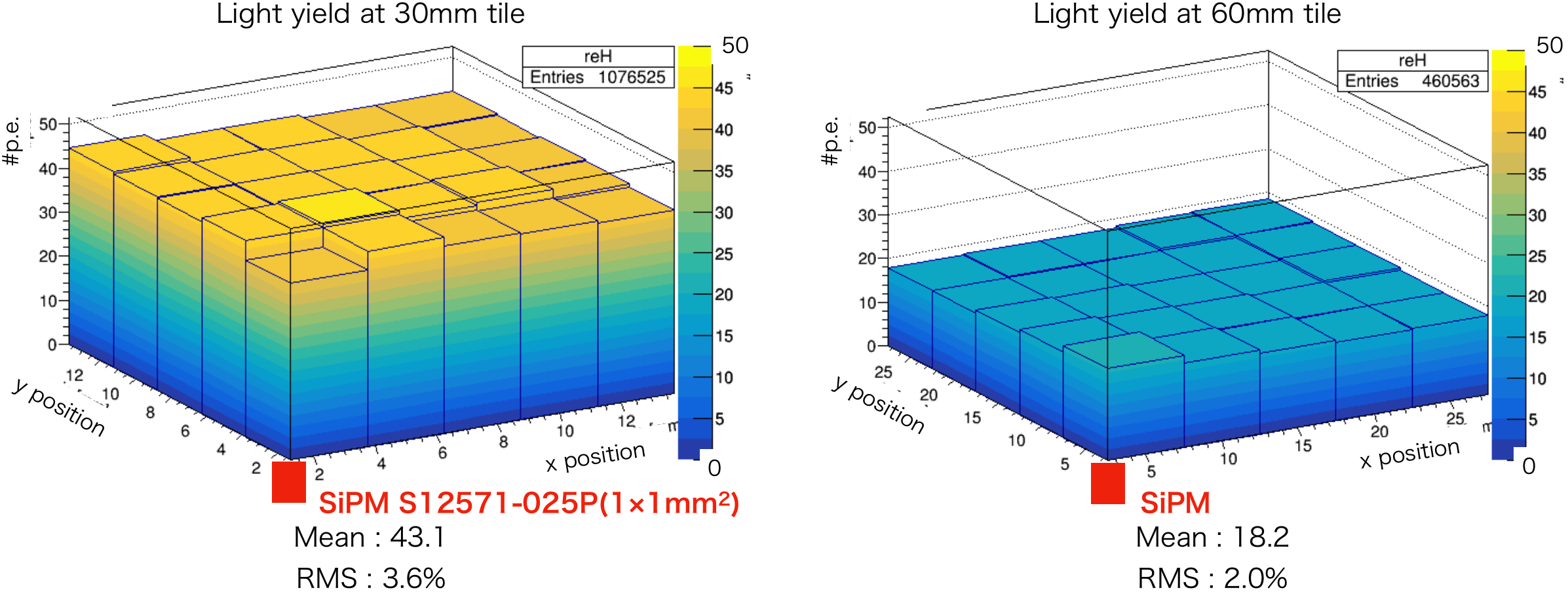}
	\caption{Light yields of 30$\times$30$~\mathrm{mm^2}$ (left) and 60$\times$60$~\mathrm{mm^2}$ (right) tiles}
	\label{fig:figA}
\end{figure}

We examined the amount and uniformity of the light yield with a prototype of 60$\times$60$~\mathrm{mm^2}$ tile. The light yield was measured to be 18.2~p.e. on average as shown in Figure \ref{fig:figA}, which is about 50\% of that of 30$\times$30$~\mathrm{mm^2}$ tile (43~p.e.). This light yield degradation can easily be recovered by using MPPCs with larger sensitive area. On the other hand, it was found that the 60$\times$60$~\mathrm{mm^2}$ tile has an excellent uniformity in tile response.

\section{Construction of detection layer with 60$\times$60$~\mathrm{mm^2}$ tiles}
\begin{figure}[H]
	\centering
	\includegraphics[width=80mm,pagebox=cropbox,clip]{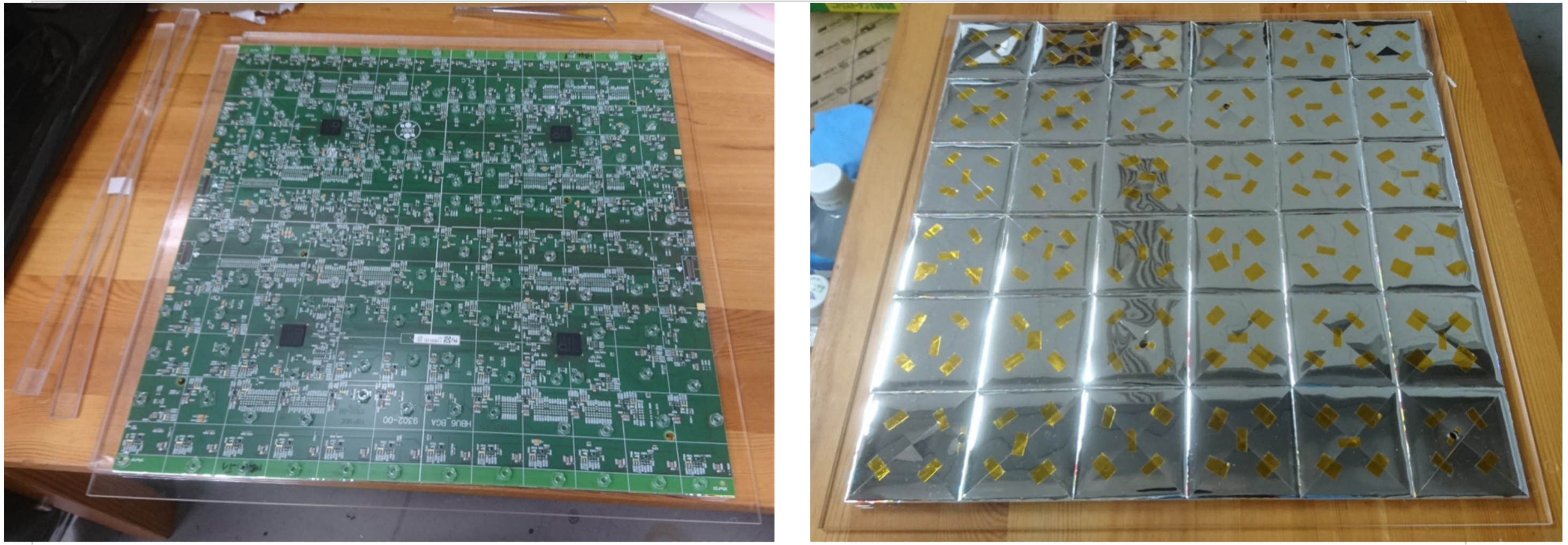}
	\caption{Completed detection layer with 60$\times$60$~\mathrm{mm^2}$ tiles. Readout electronics side (left) and scintillator tile side (right).}
	\label{fig:figB}
\end{figure}

We constructed a detection layer with 60$\times$60$~\mathrm{mm^2}$ tiles instead of standard 30$\times$30$~\mathrm{mm^2}$ tiles. The MPPC sensor size is 2$\times$2$~\mathrm{mm^2}$ instead of 1.3$\times$1.3$~\mathrm{mm^2}$ as used for the standard 30$\times$30$~\mathrm{mm^2}$ tile to compensate. Hamamatsu Photonics produced a custom MPPC with an active area of 2$\times$2$~\mathrm{mm^2}$, which is a discrete array of four pieces of TSV MPPC (S13615-1025, 1$\times$1$~\mathrm{mm^2}$). Scintillator tiles were produced by injection moulding, which is suitable for large scale production. The light yield of this polystyrene-based scintillator used for the tiles is 70\% of that of commercial PVT scintillator. Four HBUs with 144 pieces of 60$\times$60$~\mathrm{mm^2}$ tiles in total were assembled as shown in Figure \ref{fig:figB}.

\section{Installation to large prototype}

The completed layer was installed into the large prototype as the 38th layer. Almost all of  the configurations were set to be the same as standard layers, but the over voltage of the 2$\times$2$~\mathrm{mm^2}$ MPPCs were set to be lower than the standard 1.3$\times$1.3$~\mathrm{mm^2}$ MPPCs to mitigate possible saturations due to the higher hit multiplicity. The peaks of photoelectrons were still clearly resolved even with the lower over voltage.  In order to evaluate the performance, our module was tested with muon, electron and pion beam at the CERN SPS.

\begin{figure}[H]
	\centering
	\includegraphics[width=80mm,pagebox=cropbox,clip]{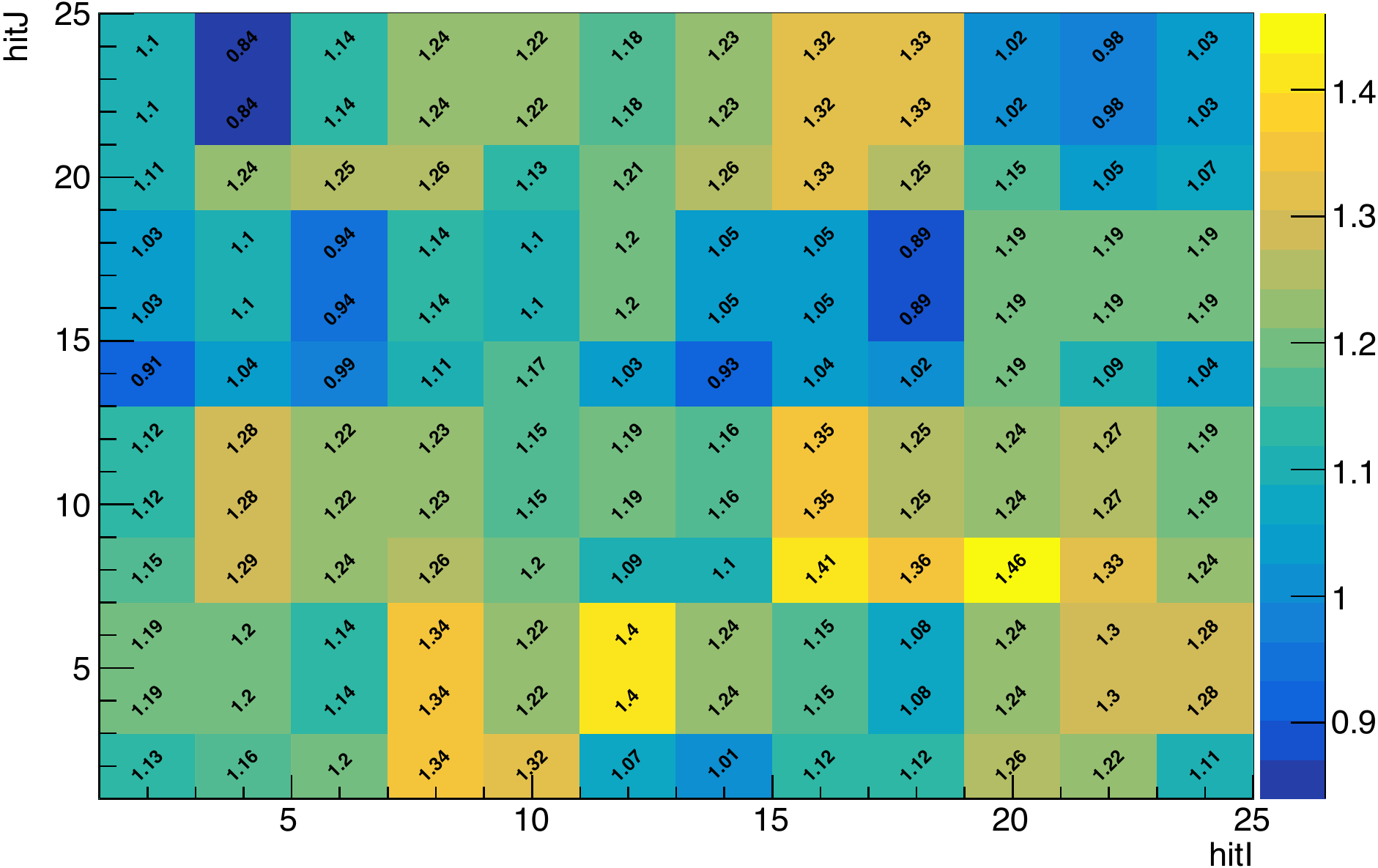}
	\caption{The relative MIP light yield for 60$\times$60$~\mathrm{mm^2}$ tiles. X and Y axes shows the channel indices.}
	\label{fig:figC}
\end{figure}

Our module worked quite well in the beam tests. Figure \ref{fig:figC} shows the measured light yield for muon as MIP, as the values relative to the average light yield for 30$\times$30$~\mathrm{mm^2}$ tile. Our module has slightly higher MIP light yield than the standard modules.

\section{Saturation}

There are two possible types of saturation; one for electronics and the other for MPPC.  The current electronics are optimized for $1.3\times1.3~\mathrm{mm^2}$ MPPCs and $30\times30\mathrm{mm^2}$ tiles, so saturation of MPPC and ADC would be an issue for $60\times60~\mathrm{mm^2}$ tiles. The $2\times2~\mathrm{mm^2}$ MPPCs have 6,400 pixels, while the $1.3\times1.3~\mathrm{mm^2}$ MPPCs have 2,700 pixels. The number of photoelectrons per MIP is the same, but the hit multiplicity can be four times higher because of the larger size. Hence, the MPPCs could be saturated more frequently at the $60\times60~\mathrm{mm^2}$ tiles. SPIROC2E, the ASIC on HBUs has high gain and low gain preamplifiers, and 12 bit ADC. A small signal is amplified at high gain, and a large signal is amplified at low gain. If the signal exceeds the maximum ADC counts (4,095), ADC saturation occurs and the output is set to 0.

\begin{figure}[H]
	\centering
	\includegraphics[width=80mm,pagebox=cropbox,clip]{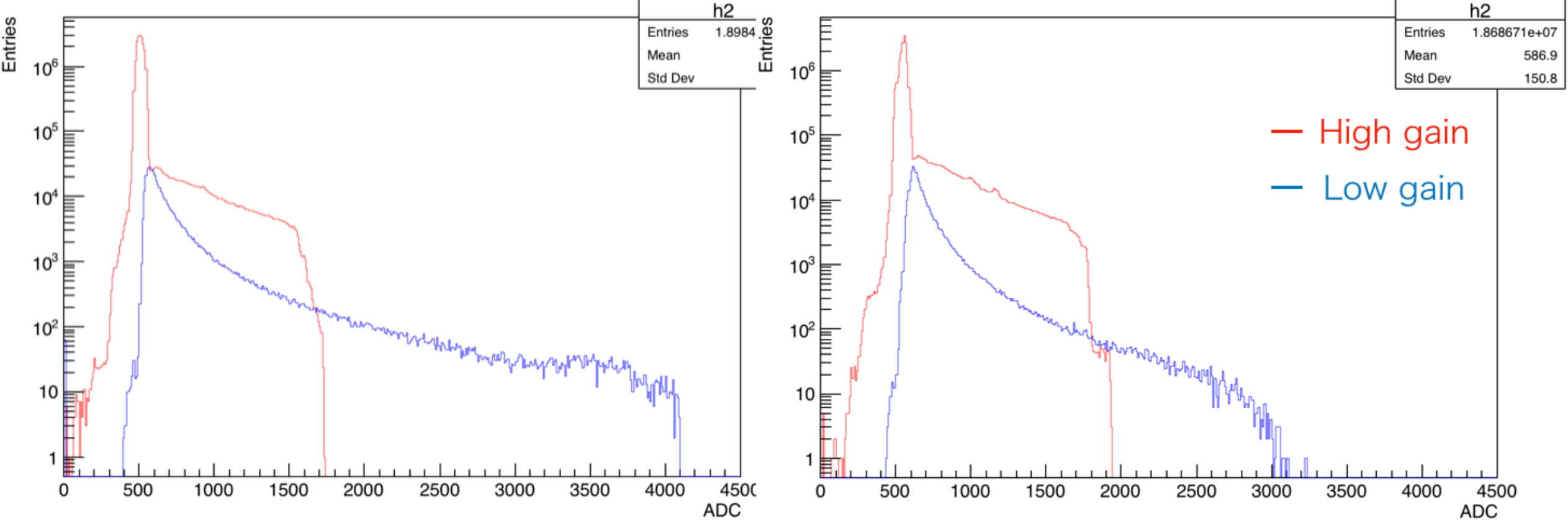}
	\caption{ADC distribution of 60$\times$60$~\mathrm{mm^2}$ tile module (left) and standard module (right) for 350~GeV pion.}
	\label{fig:figD}
\end{figure}

The MPPC and ADC saturations were observed only for the highest energy pions of 350~GeV, as shown in Figure \ref{fig:figD}. The ADCs are saturated only for 0.08\% of tile hits. The bump at 3500--4500 ADC counts is caused by the MPPC saturation. However, it should be noted that 350~GeV is much higher than the energy expected at the real experiment with $\sqrt{s}$=250--500~GeV. Therefore, these saturations would not be an issue.

\section{Summary}

As a part of granularity optimization, the mixed granularity configurations with larger scintillator tiles for outer layers of AHCAL look a viable option for cost reduction. The performance of 60$\times$60$~\mathrm{mm^2}$ tiles in terms of light yield and uniformity were demonstrated to be good enough to be used, so we constructed a detection layer with 60$\times$60$~\mathrm{mm^2}$ tiles. It was installed into the AHCAL large prototype at beam test at CERN SPS. It worked quite well with sufficient light yield. Slight saturations for ADC and MPPC were observed only at  350GeV pion, which is beyond the expected energy of particles at $\sqrt{s}$=250--500~GeV ILC.

\section{Acknowledgements}

We thank DESY FLC group as the host laboratory. 

This work supported by JSPS KAKENHI Grant Number JP17H02882 and the H2020 project AIDA-2020, GA no. 654168.

\end{multicols}

\end{document}